\documentclass[referee]{cjaa}

\usepackage{graphicx}
\input{epsf.sty}

\begin{document}
   \title{Constraining Dark Energy and Cosmological Transition Redshift with Type Ia Supernovae}

   \setcounter{page}{1}
   \author{Fa-Yin Wang
       \and Zi-Gao Dai
    \mailto{}}
   \institute{Department of Astronomy, Nanjing University, Nanjing 210093; China
             \email{dzg@nju.edu.cn}}
   \offprints{F.-Y. Wang}

\abstract{The property of dark energy and the physical reason for
acceleration of the present universe are two of the most difficult
problems in modern cosmology. The dark energy contributes about
two-thirds of the critical density of the present universe from
the observations of type-Ia supernova (SNe Ia) and anisotropy of
cosmic microwave background (CMB).The SN Ia observations also
suggest that the universe expanded from a deceleration to an
acceleration phase at some redshift, implying the existence of a
nearly uniform component of dark energy with negative pressure. We
use the ``gold'' sample containing 157 SNe Ia and two recent
well-measured additions, SNe Ia 1994ae and 1998aq to explore the
properties of dark energy and the transition redshift. For a flat
universe with the cosmological constant, we measure
$\Omega_{M}=0.28_{-0.05}^{+0.04}$, which is consistent with Riess
et al. The transition redshift is $z_{T}=0.60_{-0.08}^{+0.06}$. We
also discuss several dark energy models that define the $w(z)$ of
the parameterized equation of state of dark energy including one
parameter and two parameters ($w(z)$ being the ratio of the
pressure to energy density). Our calculations show that the
accurately calculated transition redshift varies from
$z_{T}=0.29_{-0.06}^{+0.07}$ to $z_{T}=0.60_{-0.08}^{+0.06}$
across these models. We also calculate the minimum redshift
$z_{c}$ at which the current observations need the universe to
accelerate.
 \keywords{cosmology: observations - distance scale
-supernovae: general}
   }

   \authorrunning{F. Y. Wang, Z. G. Dai}          
   \titlerunning{Constraining Dark Energy and Cosmological Transition Redshift with SNe Ia}
   \maketitle
%
%
\section{Introduction}
\label{sect:intro} Type Ia Supernovae (SNe Ia) have been
considered astronomical standard candles and used to measure the
geometry and dynamics of the universe. Kowal (1968) showed that
SNe Ia give a well-defined Hubble diagram whose intercept could
provide a good measurement of the Hubble constant. Colgate (1979)
suggested that the peak luminosity $L_{p}$ is a constant.
Subsequent observations showed that Type-I SNe should be split
(Uomoto \& Kirshier 1985; Porter \& Filippenko 1987). Theoretical
models suggested that SNe Ia arise from the thermonuclear
explosion of a carbon-oxygen white dwarf when its mass reaches the
Chandrasekhar mass (Colgate \& McKee 1969). Colgate (1979)
suggested that observations of SNe Ia around $z\simeq1$ could
measure the deceleration parameter $q_{0}$. Hansen, Jrgensen \&
N$\o$rgaard-Nielsen (1987) detected SN 1988U at $z=0.31$. At this
redshift 100 SNe Ia would have been needed to distinguish between
an open and a closed universe. Phillips (1993) discovered the
intrinsic relation in SNe Ia:
$L_{p}=a\times\bigtriangleup{m}_{15}^{b}$, where
$\bigtriangleup{m}_{15}$ is the decline rate in the optical band
15 days after the peak luminosity. This relation could be used to
explore cosmology.

 Using 16 high-redshift SNe and 34 nearby SNe, Riess
et al. (1998) found that our universe has been accelerating. Using
42 SNe Ia, Perlmutter et al. (1999) obtained the same result. SN
Ia observations also provided evidence for a decelerating universe
at redshifts higher than the transition redshift $z_{T}\simeq0.5$
(Riess et al. 2001; Turner et al. 2002; Riess et al. 2004). Tonry
et al. (2003) found that $\Omega_{M}=0.28\pm0.05$ and
$-1.48<w<-0.72$ at the 95\% confidence level for a flat universe
from high-$z$ SNe. Daly \& Djorgovski (2003) derived that the
universe changed from deceleration to acceleration at $z_{T}=0.45$
using a model-independent method. Combining the constraints from
the recent Ly-¦Á forest analysis of Sloan Digital Sky Survey
(SDSS) and the SDSS galaxy bias analysis with previous constraints
from the SDSS galaxy clustering, the latest SNe, and first-year
WMAP cosmic microwave background anisotropies, Seljak et al.
(2004) found that $\Omega_{\Lambda}=0.72\pm0.02$,
$w(z=0.3)=-0.98_{-0.12}^{+0.10}$. In the model of $w(z)=w_{0}$,
they found $w_{0}=-0.990_{-0.093-0.201-0.351}^{+0.086+0.160+0.222}
(1\sigma, 2\sigma, 3\sigma)$ . From their analysis they concluded
that the equation of state did not vary with redshift. Alam et al.
(2004) obtained the transition redshift $z_{T}=0.57\pm0.07$ from a
joint analysis of SNe Ia and CMB. Utilizing SNe Ia data, Bassett
et al. (2004) derived that the transition redshift varied from
$z_{T}=0.14$ to $z_{T}=0.59$, but Gong (2004) found
$z_{T}\simeq0.3$. Jarvis et al (2005) analysized the 75 square
degree CTIO lensing survey in conjunction with CMB and SN Ia data
and measured $w_{0}=-0.894_{-0.208}^{+0.156}(95\% confidence
level)$. When taking the dark energy model of
$w(a)=w_{0}+w_{a}(1-a)$, they found
$w_{0}=-1.19_{-1.74}^{+0.53},w_{a}=1.31_{-2.40}^{+3.04}(95\%
confidence level)$. Gong (2005) found the transition redshift was
$z_{T}\simeq0.6$ using one-parameter dark energy models. Chang et
al. (2005) gave $w_{0}=-1.29$, the deceleration parameter
$q_{0}=-0.97$ and $z_{T}=0.70$ by using the recent data of X-ray
cluster gas mass fraction. Clocchiatti et al. (2005) derived
$\Omega_M=0.79^{+0.15}_{-0.18}$ and $\Omega_\Lambda=
1.57^{+0.24}_{-0.32}$ ($1\sigma$ confidence level) if no prior
assumption is made, or $\Omega_M =0.29^{+0.06}_{-0.05}$ if
$\Omega_M + \Omega_\Lambda= 1$ is assumed, from a sample of 75
low-redshift and 47 high-redshift SNe Ia with the MLCS2k2
luminosity calibration. For a different sample of 58 low-redshift
and 48 high-redshift SNe Ia with luminosity calibrations using the
PRES method, the results were $\Omega_M=0.43^{+0.17}_{-0.19}$ and
$\Omega_\Lambda= 1.18^{+0.27}_{-0.28}$ ($1\sigma$ confidence
level) on no prior assumptions, or $\Omega_M =
0.18^{+0.05}_{-0.04}$ if $\Omega_M + \Omega_\Lambda= 1$ was
assumed. Virey et al. (2005) argued that the determi ation of the
present deceleration parameter $q_{0}$ through a simple
kinematical description could lead to wrong conclusions. A
dynamical dark energy model must be taken into account. Meng \&
Fan (2005) suggested that LAMOST redshift survey could help to
reduce the error bounds of dark energy parameters expected from
other observations. Zhang \& Wu (2005) derived a transition
redshift of $z_{T}=0.63$ using the CMB, LSS and SNe Ia data for
the holographic dark energy model.

Riess et al.(2004) selected a sample of 157 well-measured SNe Ia,
called the ``gold" sample. Assuming a flat universe, they
concluded: (1) Using the strong prior of $\Omega_{M}=0.27\pm0.04$,
fitting to a static dark energy equation of state yields
$-1.46<w<-0.78$ (95\% confidence level). (2) Assuming a possible
redshift dependence of $w(z)$ (e.g., using $w(z)=w_{0}+w_{1}z$),
the data with the strong prior indicate that the region $w_{1}< 0$
and especially the quadrant ($w_{0}>-1$ and $w_{1}<0$) are the
least favored. (3) Expand $q(z)$ into two terms:
$q(z)=q_{0}+zdq/dz$. If the transition redshift is defined through
$q(z_{T})=0$, they found $z_{T}=0.46\pm0.13$.

Currently SN Ia observations provide the most direct way to probe
the dark energy component at low redshifts. This is due to the
fact that SN data allows a direct measure of the luminosity
distance, which is directly related to the expansion law of the
universe. Since 1998, many dark energy models have been proposed
in the literature. The simplest one is that dark energy is
constant, $w(z)=w_{0}$. A linear parameterization is
$w(z)=w_{0}+w_{1}z$. Recently a simple two-parameter model
$w(z)=w_{0}+w_{1}z/(1+z)$ was discussed. By fitting the model to
SNe Ia data, $w_{0}+w_{1}>0$ was found. At high redshifts,
however, this model was not valid. In order to solve the problem,
Jassal, Bagla \& Padmanabhan (2004) modified this parameterization
to $w(z)=w_{0}+w_{1}z/(1+z)^{2}$. Hannestad \& M\"{o}rtsell (2004)
parameterized $w_{z}$ as
$w_{z}=[1+\big(\frac{1+z}{1+z_{s}}\big)^{q}]/[w_{o}^{-1}+w_{1}^{-1}(\frac{1+z}{1+z_{s}}\big)^{q}]$
. The equation of state $w(z)$ was parametrized by Lee (2005) as
$w_{z}=w_{r}\times[w_{0}\exp (qx)+\exp (qx_{c})]/[\exp (qx)+\exp
(qx_{c})]$, where $x=\ln a=-\ln (1+z)$. Johri \& Rath (2005) found
all the observational constraints are satisfied by the two above
parameterizations by the combined CMB, LSS and SN Ia data. The
Hannestad-M\"{o}rtsell model and the Lee four-parameter model for
the equation of state may be well-behaved representations of dark
energy evolution in a large range of redshifts. Here we examine
two phenomenological parametrizations for the variable dark energy
which were given by Wetterich (2004).

In our MNRAS paper, we used gamma-ray bursts and 157 SNe Ia to
constrain cosmological parameters and transition redshift in only
four dark energy models. In this paper, we systematically explore
the properties of dark energy and cosmological transition redshift
in several dark energy models. The structure of this paper is as
follows: In Section 2, we describe our analysis methods and
numerical results in a Friedmann-Robertson-Walker cosmology with
the cosmological constant. In Section 3, we present cosmological
constraints in the one-parameter dark-energy models. In Section 4,
we explore the cosmological constraints in two-parameter
dark-energy models. Conclusions and a brief discussion are
presented in Section 5.
\section{cosmology with the cosmological constant}
\label{sect:Cos} The SN Ia observations provide the currently most
direct way of probing the dark energy at low to medium redshifts
since the luminosity-distance relation is directly related to the
expansion history of the universe. The luminosity distance is
given by (Dicus \& Repko 2004)
\begin{equation}
d_{L}=\left\{
\begin{array}{l}
\displaystyle cH_{0}^{-1}(1+z)(-\Omega_{k})^{-1/2}\sin((-\Omega_{k})^{1/2}I)   \phantom{sssssssssssssssss}  \Omega_{k}<0, \\
\displaystyle cH_{0}^{-1}(1+z)I   \phantom{ssssssssssssssssssssssssssssssssssssssss}  \Omega_{k}=0,\\
\displaystyle cH_{0}^{-1}(1+z)(\Omega_{k})^{-1/2}\sinh((\Omega_{k})^{1/2}I)   \phantom{ssssssssssssssssssss}\,  \Omega_{k}>0,\\
 \end{array} \right.
\label{eqn:fc:SSC-spectrum}
\end{equation}
where
\begin{equation}
\Omega_{k}=1-\Omega_{M}-\Omega_{DE},
\end{equation}
\begin{equation}
I=\int_{0}^{z}dz/H(z),
\end{equation}
\begin{equation}
H(z)=((1+z)^{3}\Omega_{M}+f(z)\Omega_{DE}+(1+z)^{2}\Omega_{k})^{1/2},
\end{equation}
\begin{equation}
f(z)=\exp[3\int_{0}^{z}\frac{(1+w(z'))dz'}{(1+z')}],
\end{equation}
where $w(z)$ is the equation of state for dark energy and $d_{L}$
is the luminosity distance. The luminosity distance expected in a
Friedmann-Robertson-Walker (FRW) cosmology with mass density
$\Omega_{k}$ and vacuum energy density (i.e., the cosmological
constant) $\Omega_{\Lambda}$ is
\begin{eqnarray}
d_L & = & c(1+z)H_0^{-1}|\Omega_k|^{-1/2}{\rm
sinn}\{|\Omega_k|^{1/2}\nonumber \\ & & \times
\int_0^zdz[(1+z)^2(1+\Omega_Mz)-z(2+z)\Omega_\Lambda]^{-1/2}\},
\end{eqnarray}
where $\Omega_{k}=1-\Omega_{M}-\Omega_{\Lambda}$, and $\rm sinn$
is $\sinh$ for $\Omega_{k}>0$ and $\sin$ for $\Omega_{k}<0$
(Carroll et al. 1992). For $\Omega_{k}=0$, the luminosity distance
is $d_{L}=cH_{0}^{-1}(1+z)$ times the integral. With $d_{L}$ in
units of megaparsecs, the predicted distance modulus is
\begin{equation}
\mu=5\log(d_{L})+25.
\end{equation}
We can plot the Hubble diagram for the Gold sample containing 157
SNe Ia and two recent, well-measured SNe Ia 1994ae and 1998aq
(Riess et al. 2005). The likelihood functions for the parameters
$\Omega_{M}$ and $\Omega_{\Lambda}$ can be determined from
$\chi^{2}$ statistic,
\begin{equation}
\chi^{2}(H_{0},\Omega_{M},\Omega_{\Lambda})=\sum_{i=1}^{N}
\frac{[\mu_{i}(z_{i},H_{0},\Omega_{M},\Omega_{\Lambda})-\mu_{0,i}]^{2}}{\sigma_{\mu_{0,i}}^{2}+\sigma_{\nu}^{2}},
\end{equation}
where $\sigma_{\nu}$ is the dispersion in the supernova redshift
(transformed to distance modulus) due to peculiar velocities, and
$\sigma_{\mu_{0,i}}$ is the uncertainty in the individual distance
moduli. The confidence regions in the
$\Omega_{M}-\Omega_{\Lambda}$ plane can be found through
marginalizing the likelihood functions over $H_{0}$ (i.e.,
integrating the probability density $p\propto\exp^{-\chi^{2}/2}$
for all values of $H_{0}$). The Friedmann equations are
\begin{equation}
H^{2}+\frac{k}{a^{2}}=\frac{8\pi G}{3}(\rho_{M}+\rho_{r}+\rho),
\end{equation}
\begin{equation}
\dot{\rho}+3H(\rho+p)=0.
\end{equation}
The Hubble constant $H=\dot{a}/a$, the dot representing time
derivative. Here $\rho$ is defined as
\begin{equation}
\rho=\rho_{0}\exp[3\int_{0}^{z}\frac{(1+w(z'))dz'}{(1+z')}].
\end{equation}
$\rho_{M}$ is the matter energy density, $\rho_{r}$ the radiation
energy density and $z=a_{0}/a-1$ is the redshift. Combining
equations (9) and (10), we can find the acceleration equation,
\begin{equation}
\frac{\ddot{a}}{a}=-\frac{4\pi G}{3}(\rho_{M}+2\rho_{r}+\rho+3p).
\end{equation}
At $\ddot{a}=0$, the universe changes from deceleration to
acceleration phase. So we can define the transition redshift. For
the cosmological-constant model we obtain the transition redshift,
\begin{equation}
z_{T}=(\frac{2\Omega_{\Lambda}}{\Omega_{M}})^{1/3}-1.
\end{equation}
In Figure 1 we plot the Hubble diagram for the 159 SNe Ia. We use
the 159 SNe Ia data to obtain the confidence regions and
transition redshift (see Fig.\,2). For a flat universe, we obtain
$\Omega_{M}=0.28_{-0.05}^{+0.04}$. This result is consistent with
Riess et al. (2004). The best value for the transition redshift is
$z_{T}=0.60_{-0.08}^{+0.06}$. Let $z_{c}$ be the minimum redshift
at which current observations require the universe to accelerate;
it is determined from the condition $d(t_{c},t_{0})=1/H(t_{c})$.
So we have
\begin{equation}
\int_{0}^{z_{c}}\frac{dz}{\sqrt{\Omega_{M}(1+z)^{3}+\Omega_{\Lambda}(1+z)^{3(1+w)}}}=\frac{1+z_{c}}{\sqrt{{\Omega_{M}(1+z_{c})^{3}+\Omega_{\Lambda}(1+z_{c})^{3(1+w)}}}}.
\end{equation}
With a prior of $\Omega_{M}=0.27\pm0.04$, we get $z_{c}=2.01 >
z_{T}=0.60$.
\section{One-parameter dark-energy model}
\subsection{Constant parameterization}
We consider an equation of state for dark energy,
\begin{equation}
w_{z}=w_{0}.
\end{equation}
In this dark energy model, the luminosity distance for a flat
universe is (Riess et al. 2004)
\begin{equation}
d_{L}=cH_{0}^{-1}(1+z)\int_{0}^{z}dz[(1+z)^{3}\Omega_{M}+(1-\Omega_{M})(1+z)^{3(1+w_{0})}]^{-1/2}.
\end{equation}
Combining equations (11), (12) and (15), we calculate the
transition redshift through
\begin{equation}
\Omega_{M}+(1-\Omega_{M})(1+3w_{0})\times(1+z)^{3w_{0}}=0.
\end{equation}
We use the 159 SNe Ia data to obtain the confidence regions and
transition redshift, and derive $w_{0}=-0.975_{-0.15}^{+0.12}$ at
the $1\sigma$ confidence level. See Figure~3. So if we assume
$w_{0}=\rm constant$, then the SN Ia data favor $w_{0}=-1$. At the
95\% confidence level we have $-1.35<w_{0}<-0.75$. These results
are consistent with Tonry et al. (2003), Knop et al. (2003),
Bennett et al. (2003), Riess et al. (2004). The best value of the
transition redshift is $z_{T}=0.52_{-0.06}^{+0.05}(1\sigma)$. In
this dark energy model $z_{c}$ satisfies the following equation,
\begin{eqnarray}
\nonumber \\ & &
\int_{0}^{z_{c}}\frac{dz}{\sqrt{\Omega_{M}(1+z)^{3}+(1-\Omega_{M})(1+z)^{3(1+w_{0})}}}
\nonumber \\ & &
=\frac{1+z_{c}}{\sqrt{{\Omega_{M}(1+z_{c})^{3}+(1-\Omega_{M})(1+z_{c})^{3(1+w_{0})}}}}.
\end{eqnarray}
For $\Omega_{M}=0.27$ and $w_{0}=-0.975$, we get $z_{c}=2.02
> z_{T}=0.52$.

We now consider the second one-parameter dark energy equation
(Gong \& Zhang 2005),
\begin{equation}
w_{z}=\frac{w_{0}}{1+z}\exp(\frac{z}{1+z}).
\end{equation}
In this model the luminosity distance is given by
\begin{equation}
 d_{L}=cH_{0}^{-1}(1+z)\int_{0}^{z}dz[(1+z)^{3}\Omega_{M}+(1-\Omega_{M})(1+z)^{3}\exp(3w_{0}e^{\frac{z}{1+z}}-3w_{0})]^{-1/2}.
\end{equation}
Combining equations (11), (12) and (19), we can calculate the
transition redshift through
\begin{equation}
\Omega_{M}+(1-\Omega_{M})(1+\frac{3w_{0}}{1+z}e^{z/(1+z)})\times\exp(3w_{0}(e^{z/(1+z)}-1))=0.
\end{equation}
Again we use the 159 SNe Ia data to obtain confidence regions and
transition redshift and derive $w_{0}=-1.10_{-0.11}^{+0.16}$ at
the $1\sigma$ confidence level, shown in Figure~3. We obtain
$-1.32<w_{0}<-0.76$ at the 95\% confidence level.  The transition
redshift is found to be $z_{T}=0.47^{+0.07}_{-0.05}(1\sigma)$. In
this dark energy model. $z_{c}$ satisfies the following equation,
\begin{eqnarray}
\nonumber \\ & &
\int_{0}^{z_{c}}\frac{dz}{\sqrt{(1+z)^{3}\Omega_{M}+(1-\Omega_{M})(1+z)^{3}\exp(3w_{0}e^{\frac{z}{1+z}}-3w_{0})}}
\nonumber \\ & &
=\frac{1+z_{c}}{\sqrt{(1+z_{c})^{3}\Omega_{M}+(1-\Omega_{M})(1+z_{c})^{3}\exp(3w_{0}e^{\frac{z_{c}}{1+z_{c}}}-3w_{0})}}.
\end{eqnarray}
For $\Omega_{M}=0.27$ and $w_{0}=-1.10$, we get $z_{c}=1.90
> z_{T}=0.47$.

Our third one-parameter dark energy model (Gong \& Zhang 2005), is
\begin{equation}
w_{z}=\frac{w_{0}}{1+z}.
\end{equation}
Proceeding as before, we obtain the luminosity distance
\begin{equation}
 d_{L}=cH_{0}^{-1}(1+z)\int_{0}^{z}dz[(1+z)^{3}\Omega_{M}+(1-\Omega_{M})(1+z)^{3}e^{(\frac{3w_{0}z}{1+z})}]^{-1/2}.
\end{equation}
Combining equations (11), (12) and (23), we calculate the
transition redshift through
\begin{equation}
\Omega_{M}+(1-\Omega_{M})(1+\frac{3w_{0}}{1+z})\times\exp{(\frac{3w_{0}z}{1+z})}=0.
\end{equation}
Again for the 159 SNe Ia data the confidence regions and
transition redshift are obtained. We have
$w_{0}=-1.15_{-0.17}^{+0.20}$ at the $1\sigma$ confidence level
shown in Figure~3 and derive $-1.37<w_{0}<-0.78$ at the 95\%
confidence level.  The transition redshift is
$z_{T}=0.49^{+0.06}_{-0.05}(1\sigma)$. In this dark energy model
$z_{c}$ satisfies the following equation,
\begin{eqnarray}
\nonumber \\ & &
\int_{0}^{z_{c}}\frac{dz}{\sqrt{\Omega_{M}(1+z)^{3}+(1-\Omega_{M})(1+z)^{3}e^{(\frac{3w_{0}z}{1+z})}}}
\nonumber \\ & &
=\frac{1+z_{c}}{\sqrt{\Omega_{M}(1+z_{c})^{3}+(1-\Omega_{M})(1+z)^{3}e^{(\frac{3w_{0}z_{c}}{1+z_{c}})}}}.
\end{eqnarray}
For $\Omega_{M}=0.27$ and $w_{0}=-1.15$, we obtain $z_{c}=1.63
> z_{T}=0.49$.
\section{Two-parameter dark-energy model}
\subsection{Wetterich's parameterization}
In this section, we first consider the dark energy
parameterization proposed by Wetterich (Wetterich 2004):
\begin{equation}
w_{z}=\frac{w_{0}}{[1+b\ln(1+z)]^2}.
\end{equation}
In this model the luminosity distance is given by
\begin{equation}
d_{L}=cH_{0}^{-1}(1+z)\int_{0}^{z}dz[(1+z)^{3}\Omega_{M}+(1-\Omega_{M})(1+z)^{3+3w_{0}/[1+b\ln(1+z)]}]^{-1/2}.
\end{equation}
Using the above method we calculate the transition redshift
through
\begin{equation}
\Omega_{M}+(1-\Omega_{M})(1+\frac{3w_{0}}{[1+b\ln(1+z)]^2})\times(1+z)^{3+3w_{0}/[1+b\ln(1+z)]}=0.
\end{equation}
We consider a Gaussian prior of $\Omega_{M}=0.27\pm0.04$. We plot
the transition redshift probability curve. The transition redshift
is $z_{T}=0.39_{-0.05}^{+0.06}(1\sigma)$ in Figure~4, but Gong
(2004) obtained $z_{T}=0.26$, which is somewhat smaller than our
result. This may be caused by differences in the calculation
method and data.

Because the best fit for the above parameterization gives
$\Omega_{M}\sim0$ which is not physical, we apply a modified
Wetterich's parameterization (Gong 2004)
\begin{equation}
w_{z}=\frac{w_{0}}{1+b\ln(1+z)}.
\end{equation}
Combining Equations (1)--(5) and (30), the luminosity Distance is
calculated with,
\begin{equation}
d_{L}=cH_{0}^{-1}(1+z)\int_{0}^{z}dz[(1+z)^{3}\Omega_{M}+(1-\Omega_{M})(1+z)^{3}[1+b\ln(1+z)]^{3w_{0}/b}]^{-1/2}.
\end{equation}
Following the above method, we calculate the transition redshift
through
\begin{equation}
\Omega_{M}+(1-\Omega_{M})(1+\frac{3w_{0}}{[1+b\ln(1+z)]})\times[1+b\ln(1+z)]^{3w_{0}/b}=0.
\end{equation}
We use a Gaussian prior of $\Omega_{M}=0.27\pm0.04$. The
transition shift probability curve is plotted. The transition
redshift is $z_{T}=0.29_{-0.06}^{+0.07}(1\sigma)$ in Figure~4.
This result is consistent with Gong (2004).

We consider another modified Wetterich's parameterization:
\begin{equation}
w_{z}=w_{0}+\frac{w_{1}}{1+\ln(1+z)}.
\end{equation}
Combining equations (1)-(5) and (33), we can obtain the luminosity
distance with,
\begin{equation}
d_{L}=cH_{0}^{-1}(1+z)\int_{0}^{z}dz[(1+z)^{3}\Omega_{M}+(1-\Omega_{M})(1+z)^{3+3w_{0}}[1+\ln(1+z)]^{3w_{1}}]^{-1/2}.
\end{equation}
Using the above method, we calculate the transition redshift
through
\begin{equation}
\Omega_{M}+(1-\Omega_{M})(1+3w_{0}+\frac{3w_{1}}{1+\ln(1+z)})\times(1+z)^{3w_{0}}[1+\ln(1+z)]^{3w_{1}}=0.
\end{equation}
We use a Gaussian prior of $\Omega_{M}=0.27\pm0.04$. The
transition redshift probability curve is plotted. The transition
redshift is $z_{T}=0.42_{-0.07}^{+0.06}(1\sigma)$ in Figure~4, but
Gong (2004) obtained $z_{T}=0.34$, which is slightly smaller than
our result. This may be caused by differences in the calculation
method and data.
\subsection{Linder's parameterization}
The simplest parameterization including two parameters is (Maaor
et al. 2001; Weller \&Albrecht 2001; Weller \&Albrecht 2002; Riess
et al. 2004),
\begin{equation}
w_{z}=w_{0}+w_{1}z.
\end{equation}
This parameterization provides the minimum possible resolving
power to distinguish between the cosmological constant and
time-dependent dark energy. We again use the above method to
calculate the luminosity distance with,
\begin{equation}
d_{L}=cH_{0}^{-1}(1+z)\int_{0}^{z}dz[(1+z)^{3}\Omega_{M}+(1-\Omega_{M})(1+z)^{3(1+w_{0}-w_{1})}e^{3w_{1}z}]^{-1/2}.
\end{equation}
Combining equations (11), (12) and (36), we calculate the
transition redshift through,
\begin{equation}
\Omega_{M}+(1-\Omega_{M})(1+3w_{0}+3w_{1}z)\times(1+z)^{w_{0}-w_{1}}e^{3w_{1}z}=0.
\end{equation}
A Gaussian prior of $\Omega_{M}=0.27\pm0.04$ is applied here.
Using the 159 SNe Ia data to derive the confidence regions and
transition redshift, we obtain $w_{0}=-1.30_{-0.25}^{+0.18}$,
$w_{1}=1.42_{-0.83}^{+0.76}$ at the $1\sigma$ confidence level in
Figure 5. This result is consistent with Riess et al (2004). The
condition $w(0)<-1$ suggests that the dark energy is of phantom
origin. A cosmological constant lies at the $2\sigma$ confidence
level. The best value for transition redshift is
$z_{T}=0.41^{+0.06}_{-0.04}(1\sigma)$ in Figure 5. In this dark
energy model $z_{c}$ satisfies the following equation
\begin{eqnarray}
\nonumber \\ & &
\int_{0}^{z_{c}}\frac{dz}{\sqrt{\Omega_{M}(1+z)^{3}+(1-\Omega_{M})(1+z)^{3(1+w_{0}-w_{1})}e^{3w_{1}z}}}
\nonumber \\ & &
=\frac{1+z_{c}}{\sqrt{\Omega_{M}(1+z_{c})^{3}+(1-\Omega_{M})(1+z_{c})^{3(1+w_{0}-w_{1})}e^{3w_{1}z_{c}}}}.
\end{eqnarray}
For $\Omega_{M}=0.27$ , $w_{0}=-1.30$ and $w_{1}=1.42$, we obtain
$z_{c}=1.20 > z_{T}=0.41$.

The above model is not compatible with CMB data since it diverges
at high redshifts. Linder (2003) proposed an extended
parameterization which avoids this problem,
\begin{equation}
w_{z}=w_{0}+\frac{w_{1}z}{1+z}.
\end{equation}
We use again the above method to calculate the luminosity distance
with,
\begin{equation}
d_{L}=cH_{0}^{-1}(1+z)\int_{0}^{z}dz[(1+z)^{3}\Omega_{M}+(1-\Omega_{M})(1+z)^{3(1+w_{0}+w_{1})}e^{-3w_{1}z/(1+z)}]^{-1/2}.
\end{equation}
Combining equations (11), (12) and (40), we calculate the
transition redshift through
\begin{equation}
\Omega_{M}+(1-\Omega_{M})(1+3w_{0}+\frac{3w_{1}z}{1+z})\times(1+z)^{w_{0}+w_{1}}
e^{-3w_{1}z/(1+z)}=0.
\end{equation}
We obtain the confidence regions and transition redshift as
before, and obtain
$w_{0}=-1.35_{-0.28}^{+0.35}$,$w_{1}=2.02_{-1.85}^{+2.26}$ at the
$1\sigma$ confidence level in Figure 5. This result is consistent
with Riess et al. (2004). Here $w(0)<-1$ suggests that the dark
energy is of phantom origin. A cosmological constant lies at the
$2\sigma$ confidence level. We find the transition redshift to be
$z_{T}=0.31^{+0.04}_{-0.02}(1\sigma)$ in Figure~5. In this dark
energy model $z_{c}$ satisfies the following equation,
\begin{eqnarray}
\nonumber \\ & &
\int_{0}^{z_{c}}\frac{dz}{\sqrt{\Omega_{M}(1+z)^{3}+(1-\Omega_{M})(1+z)^{3(1+w_{0}+w_{1})}e^{-3w_{1}z/(1+z)}}}
\nonumber \\ & &
=\frac{1+z_{c}}{\sqrt{\Omega_{M}(1+z_{c})^{3}+(1-\Omega_{M})(1+z_{c})^{3(1+w_{0}+w_{1})}e^{-3w_{1}z_{c}/(1+z_{c})}}}.
\end{eqnarray}
For $\Omega_{M}=0.27$ , $w_{0}=-1.35$ and $w_{1}=2.02$, we obtain
$z_{c}=1.47 > z_{T}=0.31$.

By fitting the $w_{z}=w_{0}+\frac{w_{1}z}{1+z}$ model to SN Ia
data, $w_{0}+w_{1}>0$ was found, so at high redshifts this model
is not proper. In order to avoid this problem, Jassal, Bagla \&
Padmanabhan (2004) modified this parameterization to
\begin{equation}
w_{z}=w_{0}+\frac{w_{1}z}{(1+z)^2}.
\end{equation}
Proceeding as before we calculate the luminosity distance with
\begin{equation}
d_{L}=cH_{0}^{-1}(1+z)\int_{0}^{z}dz[(1+z)^{3}\Omega_{M}+(1-\Omega_{M})(1+z)^{3(1+w_{0})}e^{3w_{1}z^{2}/2(1+z)^{2}}]^{-1/2}.
\end{equation}
Combining equations (11), (12) and (44), we calculate the
transition redshift through
\begin{equation}
\Omega_{M}+(1-\Omega_{M})(1+3w_{0}+\frac{3w_{1}z}{(1+z)^{2}})\times(1+z)^{3w_{0}}e^{3w_{1}z^{2}/2(1+z)^{2}}=0.
\end{equation}
Now we consider a Gaussian prior of $\Omega_{M}=0.27\pm0.04$. We
use the 159 SNe Ia data to obtain the following confidence regions
and transition redshift. The best values are
$w_{0}=-1.50_{-0.51}^{+0.82}$ and $w_{1}=5.02_{-4.05}^{+4.86}$ at
the $1\sigma$ confidence level in Figure~5. The dark energy is
also of phantom origin because of $w(0)<-1$.  A cosmological
constant lies at the $2\sigma$ confidence level. The transition
redshift is $z_{T}=0.45^{+0.06}_{-0.05}(1\sigma)$ in Figure 5. In
this dark energy model $z_{c}$ satisfies the following equation
\begin{eqnarray}
\nonumber \\ & &
\int_{0}^{z_{c}}\frac{dz}{\sqrt{\Omega_{M}(1+z)^{3}+(1-\Omega_{M})(1+z)^{3(1+w_{0})}e^{3w_{1}z^{2}/2(1+z)^{2}}}}
\nonumber \\ & &
=\frac{1+z_{c}}{\sqrt{\Omega_{M}(1+z_{c})^{3}+(1-\Omega_{M})(1+z_{c})^{3(1+w_{0})}e^{3w_{1}z_{c}^{2}/2(1+z_{c})^{2}}}}.
\end{eqnarray}
For $\Omega_{M}=0.27\pm0.04$ , $w_{0}=-1.50_{-0.51}^{+0.82}$ and
$w_{1}=5.02_{-4.05}^{+4.86}$, we obtain $z_{c}=1.35 > z_{T}=0.45$.

\section{DISCUSSION AND CONCLUSIONS}
\label{con}In this paper we have used the Gold sample containing
157 SNe Ia plus two recently well-measured SNe Ia, 1994ae and
1998aq, to explore the property of dark energy and the transition
redshift. Our results are listed in Table 1. For a flat universe
with a cosmological constant, we measure
$\Omega_{M}=0.28_{-0.05}^{+0.04}$ and the transition redshift
$z_{T}=0.60_{-0.08}^{+0.06}$. Using accurate formulae of the
transition redshift in different dark energy models, we find that
the transition redshift varies from $z_{T}=0.29_{-0.06}^{+0.07}$
to $z_{T}=0.60_{-0.08}^{+0.06}$. The transition redshifts $z_{T}$
for all the tested parametrisations are less than that in the
$\Lambda$CDM model. From these results we can see that the
transition redshift is different in different dark energy models,
---
it is model-dependent. In these models, the dark energy properties
are consistent with a cosmological constant, so we cannot exclude
that cosmological constant acts as dark energy. We find that
$w<-1$ is more favored. For all the dark energy models, we find
$z_{c}>z_{T}$. Although there exist many dark energy models, we
are still not able to decide which model gives us the right answer
and to find out the nature of dark energy. Higher order models are
more suitable for probing the nature of dark energy and its
evolution, such as the Hannestad-M\"{o}rtsell model and Lee's
four-parameter model. However, more parameters mean more degrees
of freedom, as well as more degeneracies in the determination of
the parameters. The CMB can break degeneracies between
cosmological parameters and the SNAP mission will use a two-meter
space telescope to obtain high accuracy observations of more than
2000 SNe from $z=0.1$ to $z=1.7$. So the dark energy and the
transition redshift will hopefully be determined more accurately.
Dark energy may be a clue to new fundamental physics.

\begin{acknowledgements}
This work was supported by the National Natural Science Foundation
of China (grants 10233010 and 10221001).
\end{acknowledgements}


\newpage
\begin{table}
\begin{center}
\caption{Constraints on the cosmological parameters and transition
         redshift in several dark energy models}
\begin{tabular}{lllll}
\hline\hline%
dark energy model & $w_{0}(1\sigma)$ & $w_{1}(1\sigma)$ & $z_{T}(1\sigma)$ & $z_{c}$ \\
\hline
  $w_{z}=w_{0}$ & $-0.975_{-0.15}^{+0.12}$ & N/A & $0.52_{-0.06}^{+0.05}$ & 2.02 \\
  $w_{z}=\frac{w_{0}}{1+z}$ & $-1.15_{-0.17}^{+0.20}$ & N/A & $0.49_{-0.05}^{+0.06}$ &1.63 \\
  $w_{z}=\frac{w_{0}}{1+z}e^{z/(1+z)}$ & $-1.10_{-0.11}^{+0.16}$ & N/A & $0.47_{-0.05}^{+0.07}$ &1.90 \\
  $w_{z}=\frac{w_{0}}{1+b\ln(z)}$ & N/A & N/A & $0.29_{-0.06}^{+0.07}$ & N/A\\
  $w_{z}=\frac{w_{0}}{(1+b\ln(z))^2}$ & N/A & N/A & $0.39_{-0.05}^{+0.06}$ & N/A\\
  $w_{z}=w_{0}+\frac{w_{1}}{1+\ln(z)}$ & N/A & N/A & $0.42_{-0.07}^{+0.06}$ & N/A\\
  $w_{z}=w_{0}+w_{1}z$ & $-1.30_{-0.25}^{+0.18}$ & $1.42_{-0.83}^{+0.76}$ & $0.41_{-0.04}^{+0.06}$ &1.20\\
  $w_{z}=w_{0}+\frac{w_{1}z}{1+z}$ & $-1.35_{-0.28}^{+0.35}$ & $2.02_{-1.85}^{+2.26}$ & $0.31_{-0.02}^{+0.04}$ &1.47\\
  $w_{z}=w_{0}+\frac{w_{1}z}{(1+z)^2}$ & $-1.50_{-0.51}^{+0.82}$ & $5.02_{-4.05}^{+4.86}$ & $0.45_{-0.05}^{+0.06}$ &1.35\\
  \hline
\end{tabular}
\end{center}
\end{table}

\begin{figure}
  \includegraphics[width=135mm,height=115mm]{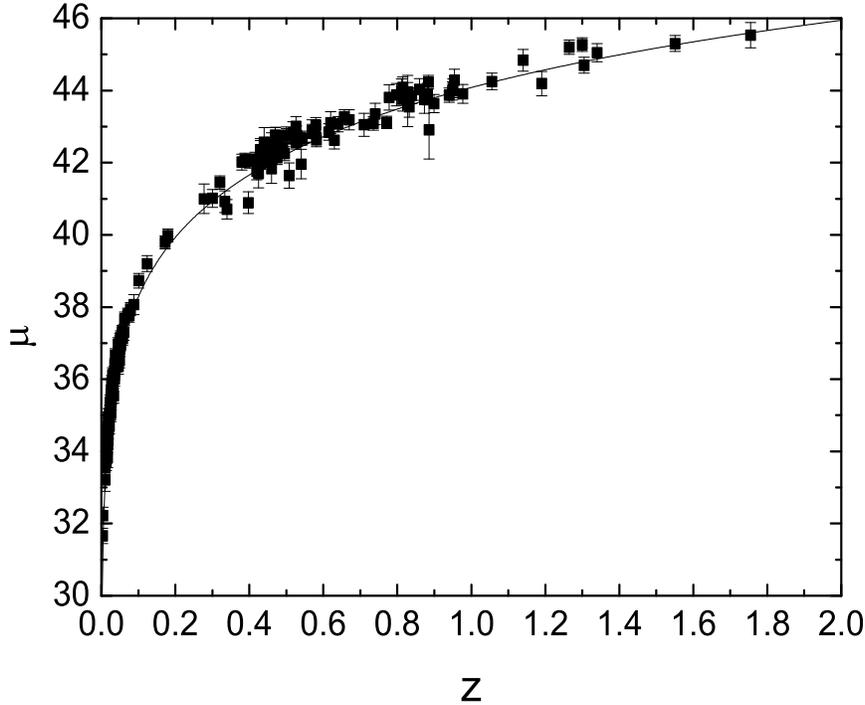}
  \caption{Hubble diagram of SNe Ia. Observed SNe Ia are shown as
dots. The solid line is the best fit for a flat cosmology:
 $\Omega_{M}=0.29$ and $\Omega_{\Lambda}=0.71$.
}
  \label{Fig1}
\end{figure}
\newpage
\begin{figure}
  \includegraphics[width=135mm,height=115mm]{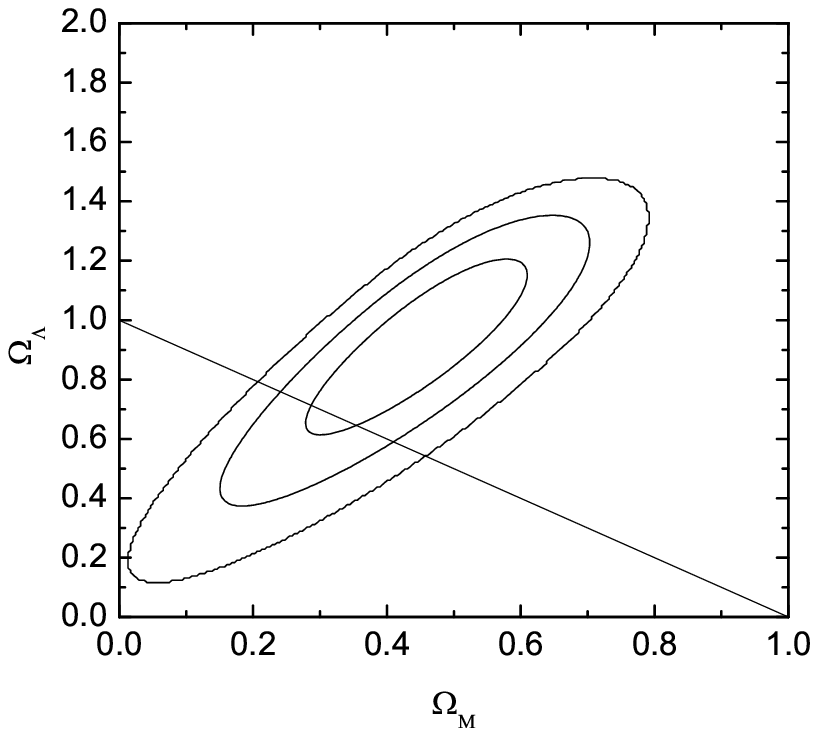}
  \includegraphics[width=135mm,height=110mm]{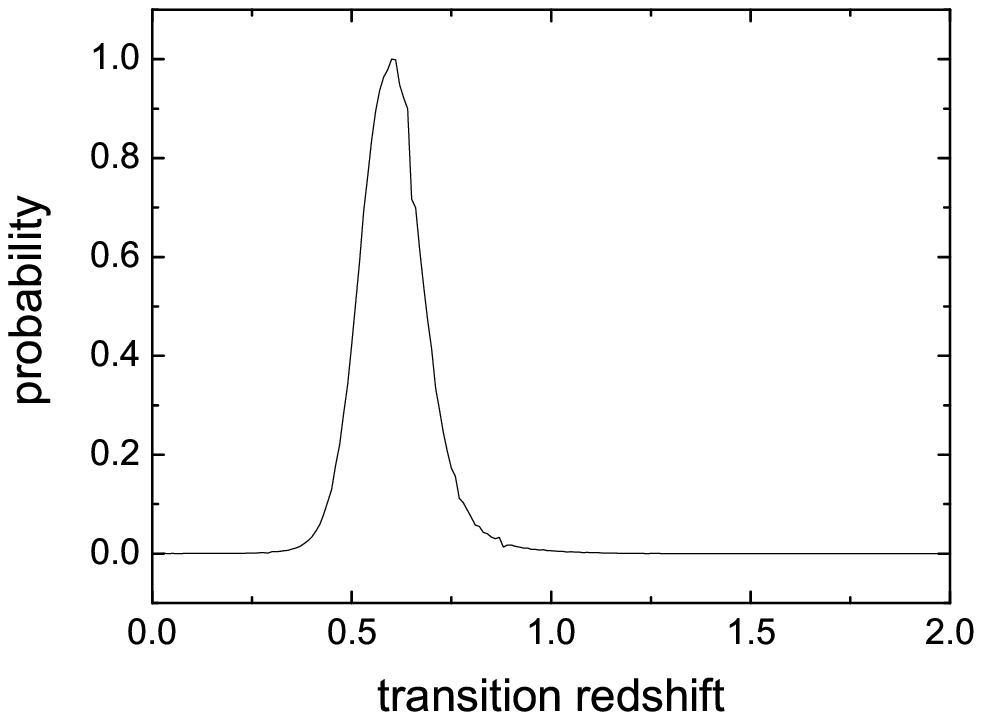}
  \caption{Left panel shows the $1\sigma$, $2\sigma$, $3\sigma$
confidence levels in the $\Omega_{M}-\Omega_{\Lambda}$ plane. The
line represents the flat universe. Right panel shows the
transition redshift probability curve. The transition redshift can
be summarily expressed by $z_{T}=0.60_{-0.08}^{+0.06}$.}
  \label{Fig2}
\end{figure}
\begin{figure}
  \includegraphics[width=135mm,height=115mm]{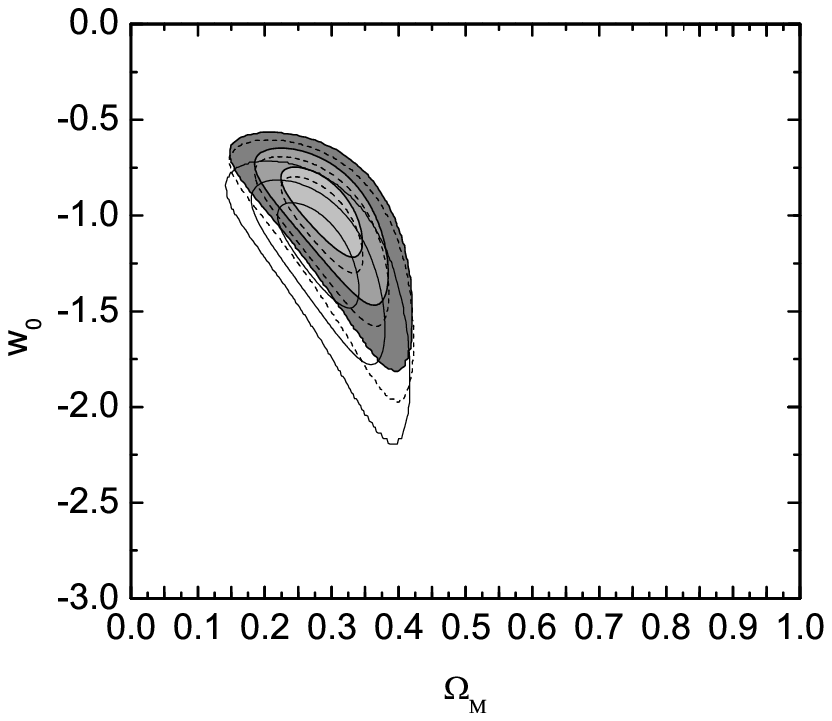}
  \includegraphics[width=135mm,height=110mm]{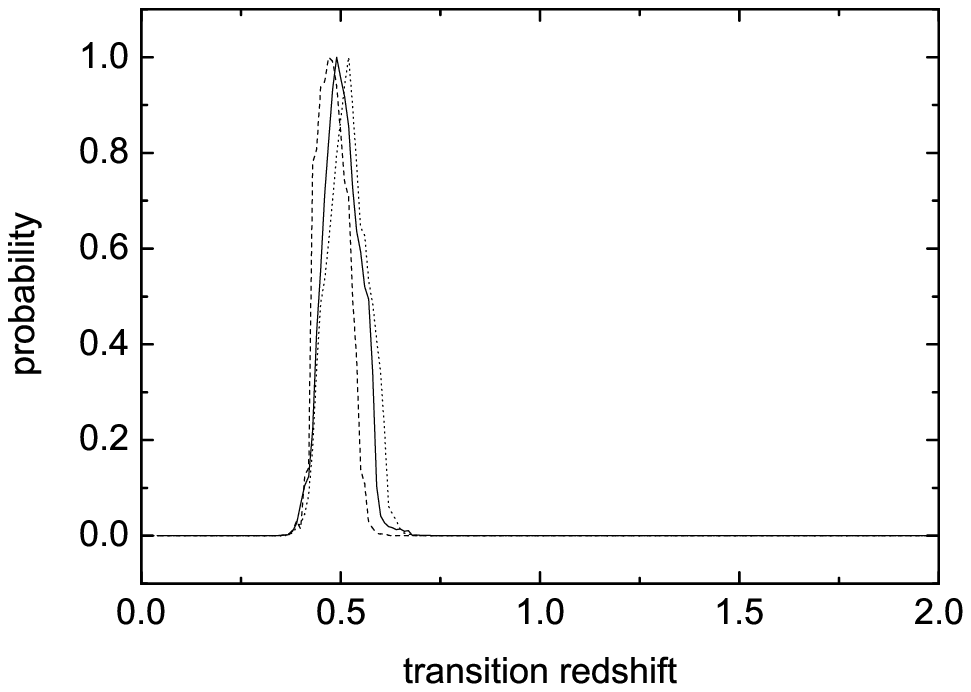}
  \caption{Left panel shows
the $1\sigma$, $2\sigma$, $3\sigma$ confidence regions in the
$\Omega_{M}-w_{0}$ plane. Grey contours refer to the $w_{z}=w_{0}$
model; dashed contours, the
$w_{z}=\frac{w_{0}}{1+z}\exp(\frac{z}{1+z})$ model; dotted
contours, to the $w_{z}=\frac{w_{0}}{1+z}$ model. Right panel
shows the transition redshift probability curve. Dotted, dashed
and full lines refer respectively to the $w_{z}=w_{0}$ model the
$w_{z}=\frac{w_{0}}{1+z}\exp(\frac{z}{1+z})$ model, and the
$w_{z}=\frac{w_{0}}{1+z}$ model.}
  \label{Fig3}
\end{figure}
\begin{figure}
  \includegraphics[width=135mm,height=115mm]{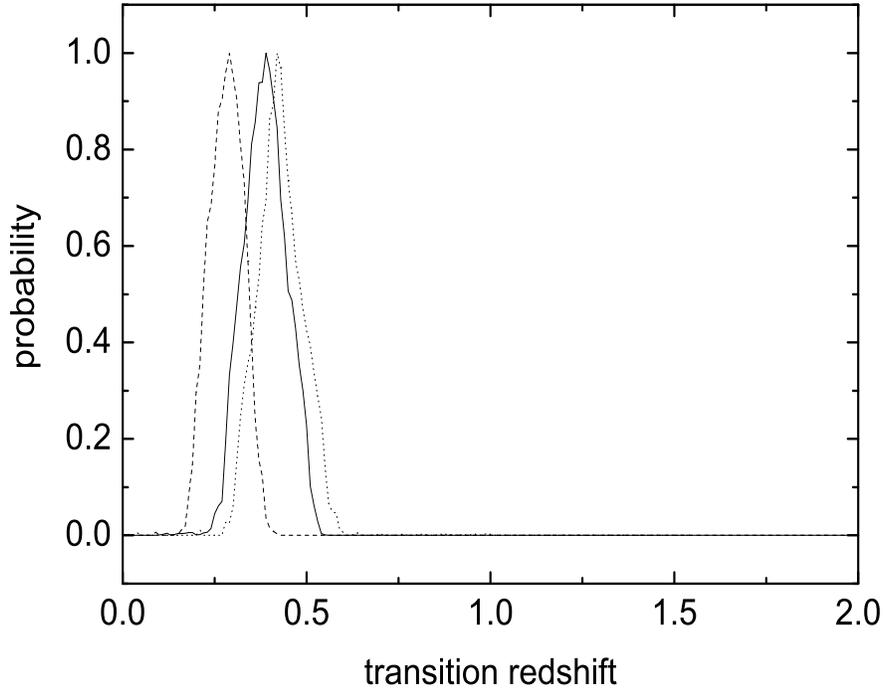}
  \caption{The transition redshift probability curve. Full line
refers to the $w_{z}=\frac{w_{0}}{(1+b\ln(z))^2}$ model; dashed
line, the $w_{z}=\frac{w_{0}}{1+b\ln(z)}$ model; dotted line, the
$w_{z}=w_{0}+\frac{w_{1}}{1+\ln(z)}$ model.}
  \label{Fig4}
\end{figure}

\begin{figure}
  \includegraphics[width=135mm,height=110mm]{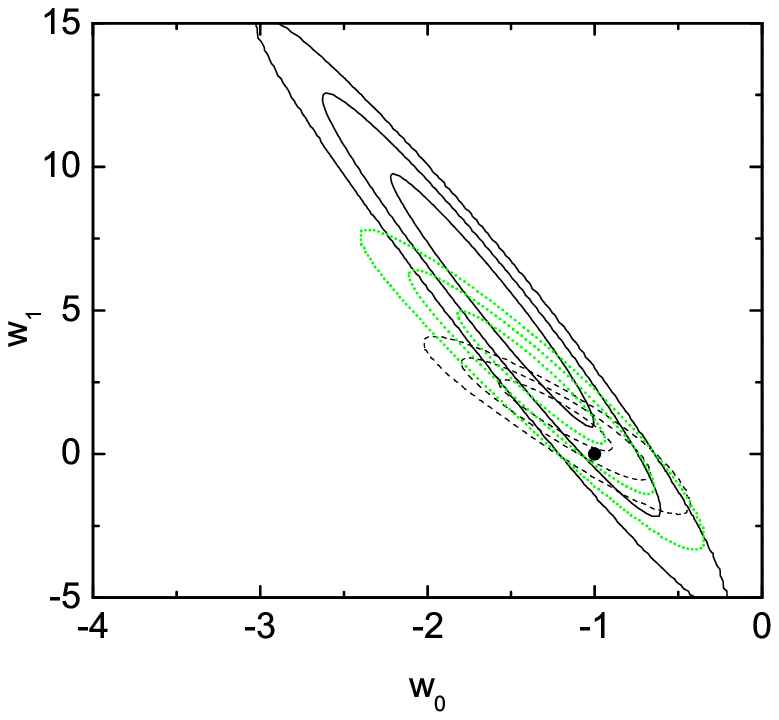}
  \includegraphics[width=135mm,height=110mm]{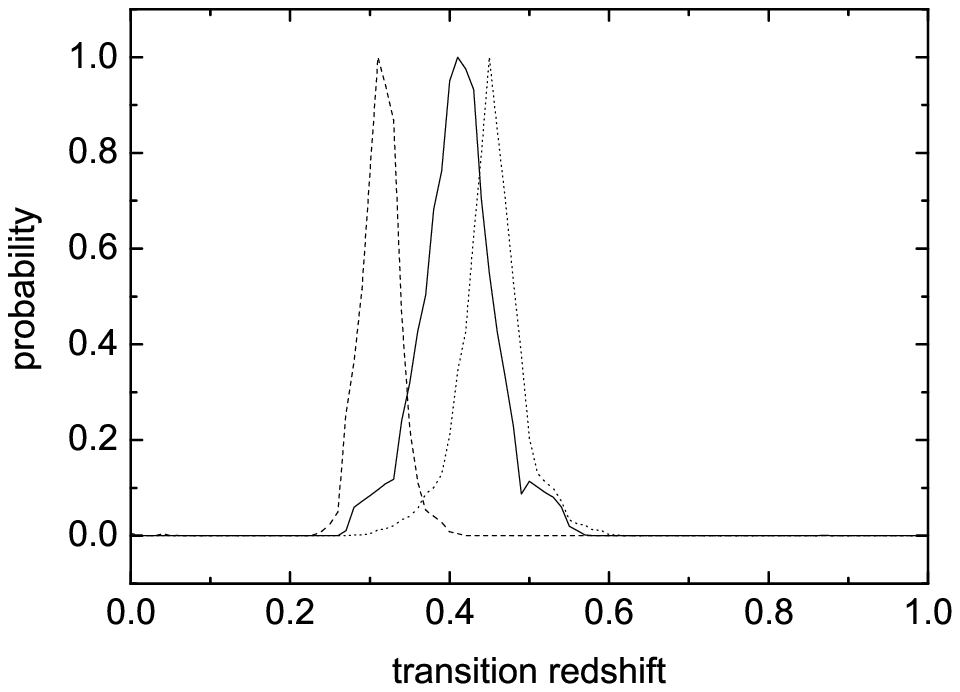}
  \caption{Left panel shows confidence regions derived from 159 SNe
Ia. Solid contours refer to the
$w(z)=w_{0}+\frac{w_{1}z}{(1+z)^{2}}$ model; green contours, the
$w(z)=w_{0}+w_{1}z/(1+z)$ model; dashed contours, the
$w_{z}=w_{0}+w_{1}z$ model. The position of a cosmological
constant, $(-1,0)$, is marked by a large dot. Right panel shows
the transition redshift probability versus. Solid line refers to
the $w_{z}=w_{0}+w_{1}z$ model; dashed line, the
$w(z)=w_{0}+w_{1}z/(1+z)$ model; dotted line, the
$w(z)=w_{0}+w_{1}z/(1+z)^{2}$ model.}
  \label{Fig5}
\end{figure}

\end{document}